\documentclass[a4paper,twocolumn,english,superscriptaddress,showpacs,aps]{revtex4}
\usepackage{amsmath}
\usepackage{amssymb}
\usepackage{graphicx}
\usepackage{color}
\usepackage{hyperref}

\begin{document}
\title{Role of uniform horizontal magnetic field on convective flow}
\author{Pinaki Pal}
\affiliation{Department of Mathematics, National Institute of Technology, Durgapur-713 209, India}
\author{Krishna Kumar}
\affiliation{Department of Physics and Meteorology, Indian Institute of Technology, Kharagpur-721 302, India} 
%\date{Received: date / Revised version: date}

\begin{abstract}
The effect of uniform magnetic field applied along a fixed horizontal direction in  Rayleigh-B\'enard convection in low-Prandtl-number fluids has been studied using a low dimensional model. The model shows the onset of convection (primary instability) in the form of two dimensional stationary rolls in the absence of magnetic field, when the Rayleigh number $R$ is raised above a critical value $R_c$. The flow becomes three dimensional at slightly higher values of Rayleigh number via wavy instability. These wavy rolls become chaotic for slightly higher values of $R$ in low-Prandtl-number ($P_r$) fluids. A uniform magnetic field along horizontal plane strongly affects all kinds of convective flows observed at higher values of $R$ in its absence. As the magnetic field is raised above certain value, it orients the convective rolls in its own direction. Although the horizontal magnetic field does not change the threshold for the primary instability, it affects the threshold for secondary (wavy) instability. It inhibits the onset of wavy instability. The critical Rayleigh number $R_o (Q,P_r)$  at the onset of wavy instability, which depends on Chandrasekhar's number $Q$ and $P_r$, increases monotonically with $Q$ for a fixed value of $P_r$. The dimensionless number $R_o (Q, P_r)/(R_c Q P_r)$ scales with $Q$ as $Q^{-1}$. A stronger magnetic field suppresses chaos and makes the flow two dimensional with roll pattern aligned along its direction.
\end{abstract}


\maketitle
%\keywords{Magnetoconvection, Instability, Chaos}  

\section{Introduction}
Convective flows in low-Prandtl-number fluids in the presence of magnetic field have been studied for many years due to its importance in geophysical and astrophysical problems~\cite{chandra,proctor}. Generally this flow is called magneto convection. There are also industrial applications of this kind of flow in crystal growth~\cite{crystal} and in fusion reactor as heat exchanger~\cite{gail}. There has been extensive theoretical and numerical studies of thermal convection in fluids in presence of external magnetic field~\cite{chandra,busse1,busse2,clever,pal1,simu1,simu2,simu3}. These studies reveal the stabilizing effect of the magnetic field on convective flow. In several experiments~\cite{expt1,expt2,expt3} it have been shown that magnetic field affects the convective flow strongly.  

Fauve et. al.~\cite{expt4,expt5} studied the effect of horizontal magnetic field, both in the longitudinal and the transverse directions, on wavy roll instability in a Rayleigh-B\`enard experiment with mercury. They found that the horizontal magnetic field inhibited the oscillatory instability and made the convection two dimensional. Libchaber et. al.~\cite{libchaber} and Hof et. al.~\cite{hof} considered the effect of external horizontal magnetic field on low-Prandtl-number thermal convection and found that the magnetic field delayed the onset of wavy instability. So magnetic field typically inhibits the oscillatory instability but there are situations when magnetic field can stimulate instabilities~\cite{mck}. Therefore, a better understanding is necessary for this kind of flow.

Numerical simulations complement experiments because very low-Prandtl-number fluids can not be achieved in the laboratory. Even the fluids like mercury ($P_r = 0.025$) do not allow very good visualization in experiments. However, the simulations are costly in terms of computer time. The simulations, due to large number of modes, often makes the understanding of the basic physics of convective flow difficult. Here, low-dimensional models play a very important roll. Low-dimensional models are useful for modeling large scale flows. It takes much less computer time and yet provide valuable information of the dynamics of the fluid flow. 

In this paper we study the effect of uniform horizontal magnetic field applied along a fixed horizontal direction ($y$-axis) using a low dimensional model. In the absence of magnetic field the model shows the onset of convection in the form of steady two dimensional rolls (2-D). The 2-D rolls become three dimensional (3-D) via wavy roll instability~\cite{busse}, when the Rayleigh number $R$ is raised slightly above critical value $R_c$. The wavy rolls
become chaotic, if $R$ is raised about $2\%$ above $R_c$ for low values of Prandtl number $P_r$. These rolls may be aligned either along $x$-axis or along $y$-axis. Two sets of rolls compete with each other with further increase in $R$. The horizontal magnetic field does not shift the threshold of the primary instability (stationary convection).
However, it delays the onset of the secondary instability in the form of wavy rolls. If Chandrasekhar's number $Q$, which is proportional to the square of the applied horizontal magnetic field ${\bf B}_0$, is raised above certain value, the rolls are oriented along the direction of the magnetic field in the horizontal plane. The dimensionless number $R_o (Q, P_r)/(R_c Q P_r)$ scales with $Q$ as $Q^{-1}$ at the onset of wavy instability. If the intensity of the magnetic field is sufficient, it suppresses chaotic flow and makes the convection periodic for relatively higher values of $R$. A stronger magnetic field suppresses the oscillatory convection even at higher values of $R$. The flow becomes steady and two dimensional in the presence of stronger horizontal magnetic field. The results of the model are also compared with those observed in experiments~\cite{expt5}.

In section II, we describe the physical system together with boundary conditions. Section III deals with the derivation of the low-dimensional model. The results of the model and their comparison with experimental and numerical simulations are described in section IV. Conclusions are given in section V. 

\section{Hydromagnetic system}
As a physical system, a thin horizontal layer of electrically conducting fluid of thickness $d$, uniform kinematic viscosity $\nu$, thermal diffusivity $\kappa$, magnetic diffusivity $\lambda$ and coefficient of volume expansion $\alpha$ is kept between two horizontal plates, and is heated uniformly from below. A uniform horizontal magnetic field ${\bf B_0} = (0,B_0,0)$ is applied along a fixed direction (say along $y$-axis). Note that we have taken $x$ and $y$-axes along horizontal directions and $z$-axis along vertical upward direction. We consider the flow of liquid metals which have magnetic Prandtl number, $P_m$ ($= \frac{\nu}{\lambda}$) of the order of $10^{-5}$~\cite{gail}. Therefore we set $P_m = 0$ for our study.
Now we choose units $d$ for length, $d^2/\nu$ for time, $\frac{B_0\nu}{\lambda}$ for the induced magnetic field and $\nu\beta d/\kappa$ for temperature, where $\beta$ is the uniform temperature gradient between the plates and get the following set of governing dimensionless equations under Boussinesq approximation: 
\begin{eqnarray}
\partial_t(\nabla^2 v_3) &=& \nabla^4 v_3 + R \nabla^2_H \theta - Q {\hat{\bf e}_3}{\bf \cdot} 
                            [{\bf \nabla\times\nabla\times}(\partial_y {\bf b})]\nonumber\\
                      & & - {\hat{\bf e}_3}{\bf \cdot\nabla\times}\left[(\mbox{\boldmath $\omega$}{\bf \cdot
                       \nabla}){\bf v}-({\bf v}{\bf \cdot \nabla})\mbox{\boldmath $\omega$}\right],\label{motion}\\
\partial_t \omega_3   &=& \nabla^2 \omega_3 + \left[(\mbox{\boldmath $\omega$}{\bf \cdot\nabla}) v_3 -({\bf v} 
                            {\bf \cdot \nabla})\omega_3\right]\nonumber\\  
                      & & + Q{\hat{\bf e}_3}{\bf \cdot}[{\bf \nabla\times}(\partial_y {\bf b})],\label{vorticity}\\
P_r[\partial_t \theta &+& ({\bf v}\cdot\nabla)\theta] = v_3 + \nabla^2\theta,\label{temp}\\
{\nabla}^2 {\bf b}    &=& - \partial_y {\bf v},\label{mag}\\
{\bf \nabla \cdot v}  &=&  {\bf \nabla \cdot b} = 0,\label{continuity}
\end{eqnarray}
where ${\bf v}$ $(x,y,z,t)$ $\equiv$ $(v_1,v_2,v_3)$ is the velocity field, ${\bf b}$ $(x,y,z,t)$ $\equiv$ $(b_1,b_2,b_3)$ the induced magnetic field due to convection, $\theta$ $(x,y,z,t)$ the  deviation in temperature field from the steady conduction profile, and $\mbox{\boldmath $\omega$}$ $\equiv$ $(\omega_1, \omega_2, \omega_3)$ $\equiv$ $\nabla\times{\bf v}$ is the vorticity field in the fluid. 
The dimensionless parameters are: Rayleigh number $R = \alpha \beta gd^4/\nu\kappa$, thermal Prandtl 
number $P_r = \nu/\kappa$ and Chandrasekhar's number $Q = \frac{B_0^2d^2}{\rho_0\lambda\nu}$, where $g$ is the acceleration due to gravity and $\rho_0$ is the reference density of the fluid. $\hat{\bf e}_3 $ is a unit vector directed vertically upward and $\nabla_H^2 = \partial_{xx} + \partial_{yy}$ is the horizontal Laplacian. 
We assume the horizontal boundaries to be stress-free, thermally and electrically good conductors. The boundary conditions are then given by

\begin{eqnarray}
\frac{\partial v_1}{\partial z} &=& \frac{\partial v_2}{\partial z} = v_3 = \theta = 0, \frac{\partial b_1}{\partial z} = \frac{\partial b_2}{\partial z} = b_3 = 0 \nonumber
\end{eqnarray}
at $z = 0, 1$. In what follows we also use one parameter $r$ called reduced Rayleigh number, being the ratio of $R$ and critical Rayleigh number $R_c = 27 {\pi}^4/4$, the threshold Rayleigh number for onset of stationary convection.

\section{Low dimensional model}
We apply standard Galerkin technique to derive a simple model to describe the convection in presence of horizontal magnetic field. The spatial dependence of the independent fields are expanded in Fourier series compatible with the boundary conditions. The expansions are truncated to describe the superposition of two mutually perpendicular sets of wavy rolls~\cite{pal} for zero ($P_r$) convection~\cite{thual} and extend the model to consider low $P_r$ convection. The expansions for vertical velocity, vertical vorticity and temperature field are as follows:
\begin{eqnarray}
v_3 &=& [W_{101}(t)\cos(k x) + W_{011}(t)\cos(k y)]\sin(\pi z)\nonumber\\
    & &+W_{112}(t)\cos(k x)\cos(k y)\sin(2 \pi z)\nonumber\\
    & &+W_{111}(t)\sin(k x)\sin(k y)\sin(\pi z)\\  
\omega_3 &=&+Z_{100}(t)\cos(k x) +Z_{010}\cos(k y)\nonumber\\
         & &+Z_{111}(t)\cos(k x)\cos(k y)\cos(\pi z)\\
\theta &=& [T_{101}(t)\cos(k x) + T_{011}(t)\cos(k y)]\sin(\pi z)\nonumber\\
    & &+[T_{112}(t)\cos(k x)\cos(k y) + T_{002}(t)]\sin(2 \pi z)\nonumber\\
    & &+T_{111}(t)\sin(k x)\sin(k y)\sin(\pi z)  
\end{eqnarray}
The horizontal components of velocity are obtained using the solenoidal property of the velocity field. In  $P_m\rightarrow 0$ limit, the magnetic field fluctuation ${\bf b}$ is slaved to ${\bf v}$, and the components of ${\bf b}$ are determined using Eq.~\ref{mag}. We then derive the model by projecting the hydrodynamical 
equations~(\ref{motion}-\ref{continuity}) on the above mentioned modes. The model, given in the Appendix, consists of twelve dimensional coupled first-order nonlinear ordinary differential equations for the above Fourier amplitudes. 

\section{Results and discussions}
We integrate the model using the ode45 solver of MATLAB to investigate the effect of magnetic field. We set $k = k_c = \pi/\sqrt{2}$. The model is  integrated in the absence of magnetic field using random initial conditions for different values of the reduced Rayleigh number $r =R/R_c$. Once the system attains steady state, magnetic field is switched on by setting $Q$ nonzero. We start our discussion by presenting the results of the model in the absence of magnetic field in the next subsection.

\subsection{Convection in the absence of magnetic field}
In the absence of magnetic field, the model shows the onset of convection in the form of steady two dimensional rolls as the value of $r$ becomes slightly above unity. As $r$ is increased above $r_o (Q,P_r)= R^{(o)}(Q, P_r)/R_c$, where
$R^{(o)}(Q, P_r)$ is the threshold for oscillatory (secondary) instability in the presence of horizontal magnetic field, the convective flow becomes three dimensional via oscillatory instability~\cite{busse}. In the model, the secondary instability appears as wavy rolls~\cite{kumar} as $r$ is raised above $r_o$. With little further increase in $r$, we find chaotically oscillating wavy rolls either oriented along $y$-axis or along $x$-axis depending upon the choice of initial conditions. These wavy rolls are observed for $1.0011 < r < 1.19$ for $P_r = .025$. For $r\ge 1.19$, the model shows chaotic competition of rolls and squares~\cite{pal,thual,kumar}. In the next subsections we present the results of the model on the effects of magnetic field on the onset of wavy roll instability and chaotic flows.

\subsection{Inhibition of oscillatory instability}
\begin{figure}[h] 
\includegraphics[height=!,width=3.5in]{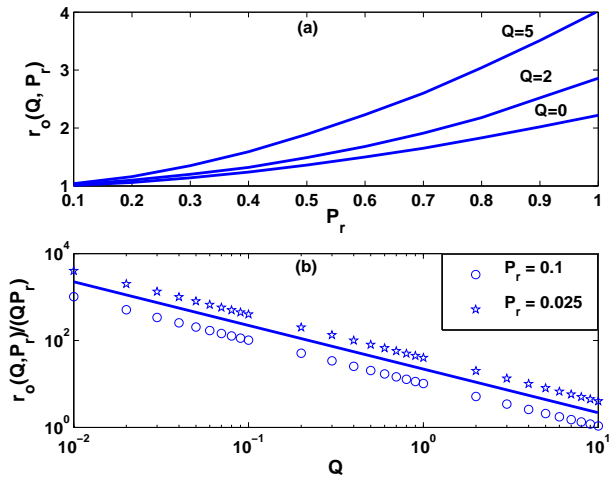}
\caption{(a)Variation of the threshold $r_o (Q,P_r)$ for wavy instability as a function of Prandtl number $P_r$  for  different values of Chandrasekhar's number $Q$. (b) Scaling of the combination $r_o (Q, P_r)/(Q P_r)$  with  $Q$ for $P_r = 0.025$ (stars) and $P_r = 0.1$ (circles). The quantity $r_o (Q, P_r)/(Q P_r)$ scales as $Q^{-1}$.  The solid line shows the same power law behaviour.}
\label{ro_Q}
\end{figure}

\begin{figure}[h] 
\includegraphics[height=!,width=3.6in]{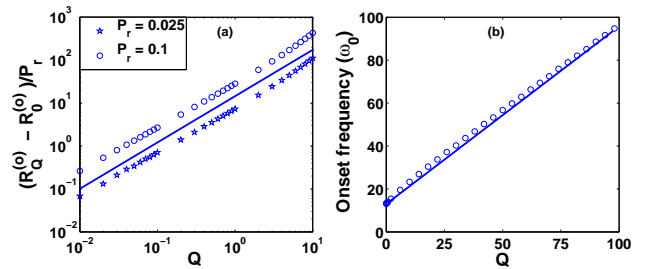}
\caption{(a)Scaling of $(R_Q^{(o)} - R_0^{(o)})/P_r$ with $Q$ for different values of $P_r$. The solid line is parallel to the best fit, and its slope is $1.1 \pm 0.01$. (b) Dimensionless frequency $\omega_o$ at the onset of oscillatory instability as a function of $Q$ for $P_r = 0.025$. The best fit (solid line) shows the linear increase of $\omega_o$ with $Q$.}
\label{scaling}
\end{figure}

The convection generated magnetic field adds a term  $-Q$ $\Delta^{-1}$ $\left(\frac{\partial^{2}{\bf v}}{\partial y^2}\right)$ only in the dimensionless Navier-Stokes equation, where $\Delta^{-1}$ denotes the inverse of Laplacian. This term represents some kind of anisotropic viscosity~\cite{expt4}. Therefore it inhibits velocity variations along the magnetic field direction. As a result, the magnetic field inhibits the onset of oscillatory instability and greatly enhances the stationary 2-D convection regime. It is observed that the onset of oscillatory instability depends both on $P_r$ and $Q$. Fig.~\ref{ro_Q}(a) clearly shows how the stability boundary of the 2-D convection regime pushed towards higher values of $r$ due to the application of uniform magnetic field in horizontal plane. This result corroborates the experimental~\cite{expt4,expt5} as well as numerical~\cite{simu1,simu2} results. The threshold for wavy instability depends on Prandtl number $P_r$ as well as Chandrasekhar's number $Q$. The combination of dimensionless numbers $r_o (Q, P_r)/(Q P_r)$ shows power law behaviour (see Fig.\ref{ro_Q}(b)), and scales with $Q$ as $Q^{-1}$. 
%%%%%%%%%%%%%%%%%%%%%%%%%%%%%%%%%%%%%%%%%%%%%%%%%%%%%%%%%%%%%%%%%%%%%%%%%%%%%%%%%%%%%%%%%%%%%%%%%%%%%%%%%%%%%%%%%
%%%%%%%%%%%%%%%%%%%%%%%%%%%%%%%%%%%%%%%%%%%%%%%%%%%%%%%%%%%%%%%%%%%%%%%%%%%%%%%%%%%%%%%%%%%%%%%%%%%%%%%%%%%%%%%%%%%%%%
We have also displayed the behaviour of the relative distance from the threshold of oscillatory instability $(R_Q^{(o)} - R_0^{(o)})/P_r$ in figure~\ref{scaling}(a) and the angular frequency $\omega_o$ at the onset of oscillatory instability in figure~\ref{scaling}(b) as a function of $Q$. Here $R_Q^{(o)}$ and $R_0^{(o)}$ are the threshold values of oscillatory (secondary) convection in the presence of the external magnetic field ($Q \neq 0$) and  in the absence of external magnetic field($Q = 0$) respectively. The results of our model is in good agreement with the numerical~\cite{busse1} and experimental~\cite{expt5} results showing the power law. The  quantity $(R_Q^{(o)} - R_0^{(o)})/P_r$ scales with $Q$ as $Q^{\alpha}$ with $\alpha =1.1$, while the experiments and direct numerical simulations gave $\alpha =1.2$. The frequency of oscillation at the onset of oscillatory instability is linearly proportional to the Chandrasekhar's number $Q$. 
%%%%%%%%%%%%%%%%%%%%%%%%%%%%%%%%%%%%%%%%%%%%%%%%%%%%%%%%%%%%%%%%%%%%%%%%%%%%%%%%%%%%%%%%%%%%%%%%%%%%%%%%%%%%%%%%%%%%%
%%%%%%%%%%%%%%%%%%%%%%%%%%%%%%%%%%%%%%%%%%%%%%%%%%%%%%%%%%%%%%%%%%%%%%%%%%%%%%%%%%%%%%%%%%%%%%%%%%%%%%%%%%%%%%%%%%%%%%

\subsection{Orientation of convective rolls}
\begin{figure}[ht]
\includegraphics[height=!,width=3.5in]{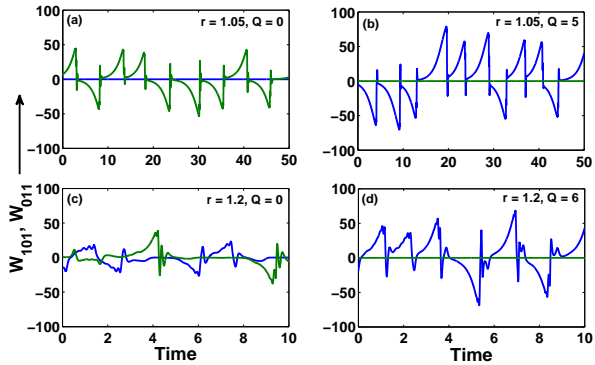}
\caption{Time series of $W_{101}$ (blue curve) and $W_{011}$ (green curve) for $P=0.025$ showing (a) chaotic wavy rolls along $x$-axis ($r=1.05$, $Q =0$) and (b) chaotic wavy rolls along $y$-axis ($r=1.05$, $Q = 5$), (c) chaotic competition of mutually perpendicular wavy rolls ($r=1.2$, $Q = 0$), and (d) chaotic wavy rolls along $y$-axis ($r =1.2$, $Q = 6$).}
\label{chaos_sq_Q}
\end{figure}

Figure~\ref{chaos_sq_Q} shows the effect of horizontal magnetic field on the orientation of the convective rolls. Fourier modes $W_{101}$ (blue curve) and $W_{011}$ (green curve) oscillate chaotically in the absence of the external magnetic field. Figure~\ref{chaos_sq_Q}(a) shows chaotic wavy rolls along $x$-axis for $r=1.05$ in the absence of external magnetic field ($Q=0$).  The set of wavy rolls initially along $x$-axis orient themselves along the direction of the magnetic field (see, figure~\ref{chaos_sq_Q}(b)) for $Q=5$. For slightly higher values of the reduced Rayleigh number ($r = 1.2$) , the model shows competition between mutually perpendicular sets of wavy rolls (Fig.~\ref{chaos_sq_Q}(c)) in the absence of external magnetic field ($Q=0$) in stead of one set of wavy rolls (Fig.~\ref{chaos_sq_Q}(a)). However, the set of wavy rolls along $x$-axis  disappears and only the rolls along $y$-axis survive (see Fig.~\ref{chaos_sq_Q}(d)) for $Q=6$. We conclude that the external magnetic field orients the  wavy rolls  along  its own direction.

\subsection{Suppression of chaos}
\begin{figure}[h]
\includegraphics[height=!,width=3.5in]{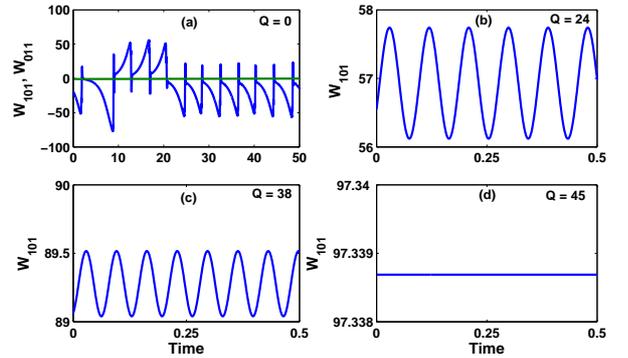}
\caption{Time series of $W_{101}$ and $W_{011}$ at $r =1.05$ and $P_r = 0.025$ for four different values of $Q$: (a) Chaotic wavy rolls along $y$-axis (for $Q = 0$) are represented by the variation of the Fourier mode $W_{101}$ (blue). The mode $W_{011}$ (green) remains zero as the rolls are already oriented along $y-$axis, (b) periodic wavy rolls ($Q = 24$), (c) periodic wavy rolls with smaller amplitude of oscillation ($Q = 38$), and (d) stationary rolls along $y$-axis ($Q = 45$).}
\label{y_roll}
\end{figure}

As mentioned above, in the absence of magnetic field the model shows three types of chaotic patterns: chaotic wavy rolls along $y$-axis, chaotic wavy rolls along $x$-axis and a competition of mutually perpendicular sets of chaotic wavy rolls depending upon the choice of initial conditions and the value of $r$. The external uniform horizontal magnetic field of moderate strength applied along $y$-axis orients the rolls along its own direction, if it is not already oriented along the magnetic field (see, figure.~\ref{chaos_sq_Q}). Once the roll-patterns are oriented along the field direction, the effect of further increase of magnetic field is similar to its effect on a roll-patterns already oriented along $y$-axis.  Figure~\ref{y_roll}(a) displays the time series of the Fourier modes $W_{101}$ (blue) and $W_{011}$ (green) in the absence of magnetic field ($Q = 0$). This represents chaotic wavy rolls along $y$-axis. If the external magnetic field along $y$-axis is gradually increased to a large value, chaotic flow is suppressed and a periodic flow develops.
The time series of the Fourier mode $W_{101}$ corresponding to the periodic flow is displayed in figure~\ref{y_roll}(b). Further increase in the field strength makes the amplitude of oscillation of the mode $W_{101}$ smaller (see, figure~\ref{y_roll}(c)). The flow becomes stationary and two-dimensional (2-D) as field strength is increased more. Figure~\ref{y_roll}(d) shows the time series of $W_{101}$ corresponding to 2-D stationary solution.
\begin{figure}[h]
\includegraphics[height=!,width=3.5in]{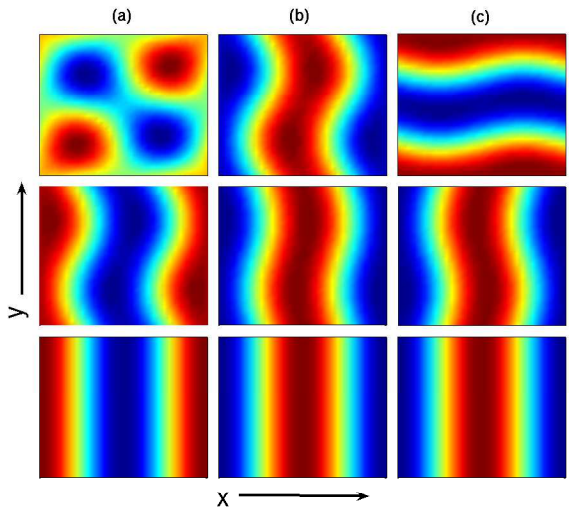}
\caption{Isotherms at $z = 0.5$ for different values of $r$ and $Q$. (a) The left column displays isotherms for $r = 1.2$, and for three values of $Q$. The chaotic competition of rolls and squares (top row) for $Q=0$, time-periodic wavy rolls (middle row) along $y$-axis for $Q=60$, and stationary 2-D rolls (bottom row) for $Q=80$.  (b) The middle column displays isotherms for $r=1.05$ and three different values of $Q$. Chaotic wavy rolls (top) along $y$-axis for $Q=0$ become time-periodic wavy rolls (middle) for $Q=24$ and finally spatially periodic stationary 2-D rolls (bottom) along $y$-axis for $Q=45$. (c) Isotherms for $r=1.05$ but with initial conditions different than used in (b) are displayed in right column. Chaotic wavy rolls along $x$-axis (top) orient themselves along $y$-axis as time-periodic wavy rolls (middle) for $Q=38$, and finally become 2-D stationary rolls along $y$-axis for $Q=45$.}
\label{pattern}
\end{figure}
The mechanism of the suppression of chaos and reorientation of wavy rolls  in the direction of the external magnetic field ($y$-axis) is displayed very clearly in figure~\ref{pattern}. It displays the isotherms at $z = 0.5$  computed from the model for different values $r$ and $Q$. The top row shows the various convective patterns in the absence of magnetic field ($Q=0$): (a) chaotic square patterns ($r= 1.2$), (b) chaotic wavy rolls along $y$-axis ($r=1.05$) and (c) chaotic wavy rolls along $x$-axis ($ r=1.05$). The effect of increasing magnetic field on different patterns are shown in three columns of figure~\ref{pattern}. As $Q$ increases,  the wavy rolls orient themselves along the direction of the horizontal magnetic field. The chaos is suppressed for relatively higher values of $Q$ and the flow becomes steady and two dimensional for further higher values of $Q$ .
\begin{figure}[h]
\includegraphics[height=!,width=3.5in]{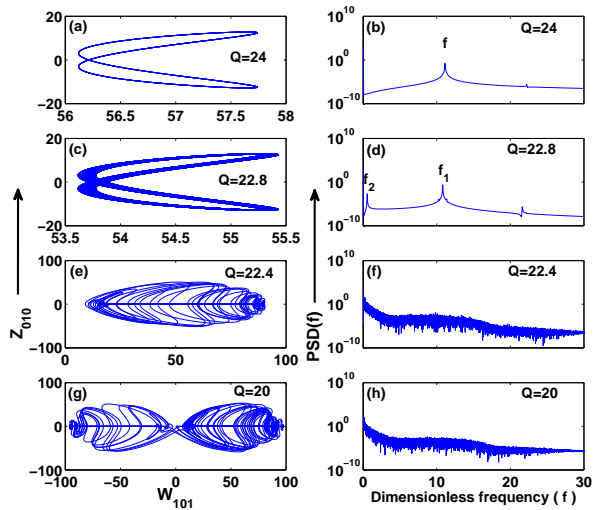}
\caption{Quasi-periodic route to chaos as a function of $Q$ for $P_r = 0.025$ and $r=1.05$. The left column displays projection of the phase space trajectories on $W_{101} - Z_{010}$ plane for four different values of $Q$. The right column shows power spectral density (PSD) of the Fourier mode $W_{101}$ as a function of the dimensionless frequency $f$ for the corresponding values of $Q$.  Time-periodic behaviour of the convective flow (a) for $Q=24$ and one peak in PSD corresponding to the periodic flow (b). The temporal quasi-periodic behaviour of the flow (c) for $Q=22.8$ and two peaks (d) confirming the quasi-periodic flow. Chaotic flow (e) for $Q=22.4$ and its PSD (f). The chaotic flow becomes  more complex for $Q=20$ (g) with flattening of peaks in PSD (h).}
\label{route_to_chaos}
\end{figure}
%%%%%%%%%%%%%%%%%%%%%%%%%%%%%%%%%%%%%%%%%%%%%%%%%%%%%%%%%%%%%%%%%%%%%%%%%%%%%%%%%%%%%%%%%%%%%%%%%%%%%%%%%%%%%%%%%%%%
The route to chaotic flow from stationary convective motion is displayed in figure~\ref{route_to_chaos}. As Chandrasekhar's number $Q$ is decreased from higher to lower value, the system shows a series of bifurcations. For much higher values of $Q$ the convection is stationary in the form of 2-D rolls along the direction of the magnetic field. With further decrease in $Q$, there is a Hopf bifurcation from stationary state. The convective flow shows time periodic behaviour (figure~\ref{route_to_chaos}a-b). As $Q$ is lowered below certain value of $Q$, one more independent frequency develops in the flow. Consequently, the convection becomes quasi-periodic in time (figure~\ref{route_to_chaos}c-d). Further decrease of $Q$ leads to chaotic motion (figure~\ref{route_to_chaos}e-f).
The route to chaos is via temporal quasi-periodic convection. The chaotic motion becomes more complicated with further decrease in $Q$ (see figure~\ref{route_to_chaos}g-h).
%%%%%%%%%%%%%%%%%%%%%%%%%%%%%%%%%%%%%%%%%%%%%%%%%%%%%%%%%%%%%%%%%%%%%%%%%%%%%%%%%%%%%%%%%%%%%%%%%%%%%%%%%%%%%%%%%%%

\begin{figure}[h]
\includegraphics[height=!,width=3.5in]{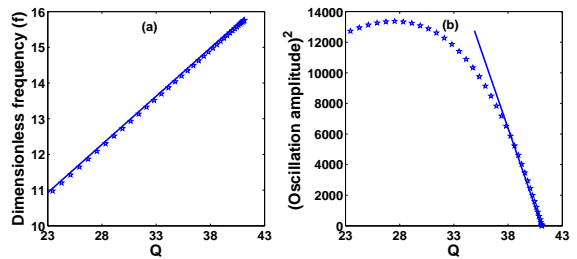}
\caption{(a) Nondimensional frequency $f$ of the wavy convection with $Q$ at a fixed value of the reduced Rayleigh number $r=1.05$. The best fit (solid line) shows the  linear dependence of $f$ with $Q$. (b) Square of the amplitude of the periodic wavy convection as a function of $Q$ at $r = 1.05$.}
\label{fre_amp}
\end{figure}

The amplitude and frequency of oscillatory convection due to Hopf bifurcation, as $Q$ is lowered from higher to lower values for a fixed value of $r$, are displayed in figure~\ref{fre_amp}. The dimensionless frequency decreases linearly with the decrease in $Q$~(figure~\ref{fre_amp}a). The square of the amplitude of the oscillatory motion increases linearly with the decrease in $Q$ near the onset of Hopf bifurcation~(figure~\ref{fre_amp}b). However, the square of the amplitude starts increasing slower than linear with decrease in $Q$ away from the onset of oscillatory behaviour. Ultimately, its growth is stopped and it starts decreasing with decreasing $Q$.  This may be the influence of quasi-periodic bifurcation present in the model. 
%%%%%%%%%%%%%%%%%%%%%%%%%%%%%%%%%%%%%%%%%%%%%%%%%%%%%%%%%%%%%%%%%%%%%%%%%%%%%%%%%%%%%%%%%%%%%%%%%%%%%%%%%%%%%%%%%%%%%%
%%%%%%%%%%%%%%%%%%%%%%%%%%%%%%%%%%%%%%%%%%%%%%%%%%%%%%%%%%%%%%%%%%%%%%%%%%%%%%%%%%%%%%%%%%%%%%%%%%%%%%%%%%%%%%%%%%%

\begin{figure}[h] 
\includegraphics[height=!,width=3.5in]{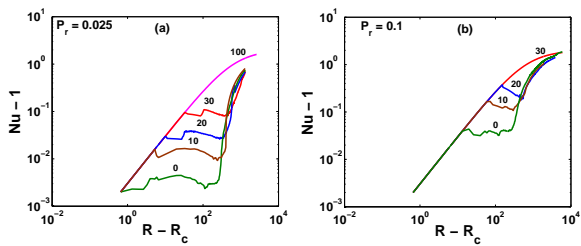}
\caption{Plot of convective heat flux ($Nu -1$) as a function of relative Rayleigh number ($R-R_c$) for  $P_r = 0.025$ (a) and $P_r = 0.1$ (b) for several values of $Q$.}
\label{nusselt}
\end{figure}
%%%%%%%%%%%%%%%%%%%%%%%%%%%%%%%%%%%%%%%%%%%%%%%%%%%%%%%%%%%%%%%%%%%%%%%%%%%%%%%%%%%%%%%%%%%%%%%%%%%%%%%%%%%%%%%%%%%%%%
We have also plotted the Nusselt number obtained from the model as a function of Rayleigh number $R$ in figure~\ref{nusselt}. The time averaged Nusselt number at the onset of convection has strong similarity with the behaviour of Nusselt number computed in the direct numerical simulations~\cite{busse2}. In addition, the results of the model also show the convergence of Nusselt numbers for higher values of $R$ as observed in simulation~\cite{busse2} for various values of $Q$. For intermediate values of $R$, the Nusselt number obtained from the model shows differences with the same obtained in DNS. This may be due to severe truncation of the Fourier expansion of various fields in the model.

\section{Conclusions}
In this paper we presented a low-dimensional model to study the effect of uniform horizontal magnetic field on Rayleigh-B\'enard convection in low-Prandtl-number fluids. The model is capable of capturing convective patterns in the form of two sets of mutually perpendicular straight (2-D) rolls as well wavy (3-D) rolls. The external field affects the flow patterns significantly. It inhibits the onset of wavy rolls, which appear at the secondary instability. The stationary straight rolls at primary instability as well as wavy rolls at secondary instability orient themselves along the direction of the magnetic field, if it is sufficiently strong. The threshold value of the Rayleigh number at the onset of wavy instability shows power law behaviour. Stronger magnetic field causes the suppression of chaotic flow. The results obtained from this simple model have good qualitative agreement with those observed in experiments and simulations.\\

PP thanks M. K. Verma, IIT Kanpur, India, for useful discussions.

\begin{center}
{\bf Appendix : The model}
\begin{eqnarray}
\dot{{\bf U}} &=& -\frac{1}{6}(9\pi^2 {\Bbb I} + 2Q{\Bbb A}){\bf U} + \frac{9\pi^4}{4} r {\bf T} + \frac{\pi}{4} V_2 {\Bbb B} {\bf U}\nonumber\\
                 & &+\frac{1}{6\pi}(\pi V_1 {\Bbb C} - 2X){\bf Z}, \nonumber\\
\dot{V_1} &=& -\frac{1}{4}(8\pi^2 + Q)V_1 + \frac{27}{8}\pi^4 r S_1 - {\Bbb C}{\bf U}\cdot{\bf Z},\nonumber\\
\dot{V_2} &=& - \frac{1}{10}(50\pi^2 + Q)V_2 + \frac{27}{20}\pi^4 r S_2 - \frac{3\pi}{5}U_1U_2,\nonumber\\
\dot{{\bf Z}} &=& -\frac{1}{2}(\pi^2 {\Bbb I} + 2Q{\Bbb D}){\bf Z} + \frac{\pi^2}{8}V_1{\Bbb C}{\bf U} + \frac{\pi}{8}X{\bf U},\nonumber\\
\dot{X} &=& -\frac{1}{4}(8\pi^2 + Q)X + \pi {\bf U}\cdot{\bf Z},\nonumber\\
\dot{{\bf T}} &=& -\frac{3\pi^2}{2P_r}{\bf T} + \frac{1}{P_r} {\bf U} + \frac{1}{2}S_1 {\bf Z} +\frac{\pi}{4}(S_2{\Bbb B} + 4 {\Bbb I} Y){\bf U},\nonumber\\
\dot{S_1} &=& -\frac{2\pi^2}{P_r}S_1 + \frac{1}{P_r}V_1 - {\Bbb C}{\bf Z}\cdot {\bf T} + \pi V_1 Y,\nonumber\\
\dot{S_2} &=& -\frac{5\pi^2}{P_r}S_2 + \frac{1}{P_r}V_2 - \frac{\pi}{2}{\bf U}\cdot{\Bbb B}{\bf T},\nonumber\\
\dot{Y} &=& -\frac{4\pi^2}{P_r}Y - \frac{\pi}{2}{\bf U}\cdot{\bf T} - \frac{\pi}{4}V_1S_1,\nonumber
\end{eqnarray}
\end{center}
where ${\bf U} = (U_1, U_2)^T = (W_{101}, W_{011})^T$, $(V_1, V_2) = (W_{111}, W_{112})$,                      ${\bf Z} = (Z_1, Z_2)^T = (Z_{010}, Z_{100})^T$, ${\bf T} = (T_1, T_2)^T = (T_{101}, T_{011})^T$, $(S_1, S_2)          = (T_{111}, T_{112})$,  $X = Z_{111}$,  $Y = T_{002}$,                                                          ${\Bbb A} = \left(\begin{array}{c} 0~~0\\0~~1\end{array}\right)$,                                             ${\Bbb B} = \left(\begin{array}{c} 0~~1\\1~~0\end{array}\right)$,                                             ${\Bbb C} = \left(\begin{array}{c} 1~~~~0\\0~~-1\end{array}\right)$,  
${\Bbb D} = \left(\begin{array}{c} 1~~0\\0~~0\end{array}\right)$, and ${\Bbb I} = \left(\begin{array}{c} 1~~0\\0~~1\end{array}\right)$. The dot ($\cdot$) operation implies standard inner product, and the superscript $T$ denotes Transpose of a matrix.

\end{document}